\documentclass[
preprint,
superscriptaddress,
amsmath,amssymb,
aps, physrev,
]{revtex4-2}

\usepackage{graphicx}
\usepackage{dcolumn}
\usepackage{bm}
\usepackage{hyperref}
\usepackage{siunitx}
\DeclareSIUnit\bar{bar}
\usepackage{wasysym}

\newcommand{\SLAC}{SLAC National Accelerator Laboratory, Menlo Park, California 94025, USA}
\newcommand{\StanfordApplPhys}{Department of Applied Physics, Stanford University, Stanford, California 94305, USA}
\newcommand{\TUDA}{Institut f{\"u}r Kernphysik, Technische Universit{\"a}t Darmstadt, Darmstadt, Germany}
\newcommand{\LLNL}{Lawrence Livermore National Laboratory, Livermore, California 94550, USA}
\newcommand{\CLF}{Central Laser Facility, STFC Rutherford Appleton Laboratory, Didcot, UK}
\newcommand{\QUB}{School of Mathematics and Physics, Queen’s University Belfast, Belfast, UK}
\newcommand{\Alberta}{Department of Electrical and Computer Engineering, University of Alberta, Edmonton, AB, T6G1H9, Canada}
\newcommand{\Michigan}{G\'erard Mourou Center for Ultrafast Optical Science, University of Michigan, Ann Arbor, MI 48109-2099, USA}
\newcommand{\Imperial}{The John Adams Institute for Accelerator Science, Imperial College London, London, SW7 2AZ, UK}
\newcommand{\ELI}{ELI Beamlines Facility, The Extreme Light Infrastructure ERIC, Za Radnic\'{i} 835, 25341 Doln\'{i} B\v{r}e\v{z}any, Czech Republic}
\newcommand{\Strath}{Department of Physics, SUPA, University of Strathclyde, Glasgow G4 0NG, UK}
\newcommand{\Cockcroft}{The Cockcroft Institute, Sci-Tech Daresbury, Warrington, WA4 4AD, UK}
\newcommand{\CTU}{Faculty of Nuclear Sciences and Physical Engineering, Czech Technical University in Prague, Prague,
Czech Republic}
\newcommand{\StanfordMech}{Department of Mechanical Engineering, Stanford University, Stanford, California 94305, USA}

\begin{document}

\title{Characterization and automated optimization of laser-driven proton beams from converging liquid sheet jet targets}

\author{G. D. Glenn}
\email{gdglenn@slac.stanford.edu}
\affiliation{\SLAC}
\affiliation{\StanfordApplPhys}
\author{F. Treffert}
\affiliation{\SLAC}
\affiliation{\TUDA}
\affiliation{\LLNL}
\author{H. Ahmed}
\author{S. Astbury}
\affiliation{\CLF}
\author{M. Borghesi}
\affiliation{\QUB}
\author{N. Bourgeois}
\affiliation{\CLF}
\author{C. B. Curry}
\affiliation{\SLAC}
\affiliation{\Alberta}
\author{S. J. D. Dann}
\affiliation{\CLF}
\author{S. DiIorio}
\affiliation{\Michigan}
\author{N. P. Dover}
\affiliation{\Imperial}
\author{T. Dzelzainis}
\affiliation{\CLF}
\author{O. Ettlinger}
\affiliation{\Imperial}
\author{M. Gauthier}
\affiliation{\SLAC}
\author{L. Giuffrida}
\affiliation{\ELI}
\author{R. J. Gray}
\affiliation{\Strath}
\affiliation{\Cockcroft}
\author{J. S. Green}
\affiliation{\CLF}
\author{G. S. Hicks}
\affiliation{\Imperial}
\author{C. Hyland}
\affiliation{\QUB}
\author{V. Istokskaia}
\affiliation{\ELI}
\affiliation{\CTU}
\author{M. King}
\affiliation{\Strath}
\affiliation{\Cockcroft}
\author{B. Loughran}
\affiliation{\QUB}
\author{D. Margarone}
\affiliation{\QUB}
\affiliation{\ELI}
\author{O. McCusker}
\affiliation{\QUB}
\author{P. McKenna}
\affiliation{\Strath}
\affiliation{\Cockcroft}
\author{Z. Najmudin}
\affiliation{\Imperial}
\author{C. Parisua\~na}
\affiliation{\SLAC}
\affiliation{\StanfordMech}
\author{P. Parsons}
\affiliation{\QUB}
\author{C. Spindloe}
\affiliation{\CLF}
\author{M. J. V. Streeter}
\affiliation{\QUB}
\author{D. R. Symes}
\affiliation{\CLF}
\author{A. G. R. Thomas}
\affiliation{\Michigan}
\author{N. Xu}
\affiliation{\Imperial}
\author{S. H. Glenzer}
\email{glenzer@slac.stanford.edu}
\affiliation{\SLAC}
\author{C. A. J. Palmer}
\email{c.palmer@qub.ac.uk}
\affiliation{\QUB}

\date{\today}

\begin{abstract}
Compact, stable, and versatile laser-driven ion sources hold great promise for applications ranging from medicine to materials science and fundamental physics. While single-shot sources have demonstrated favorable beam properties, including the peak fluxes necessary for several applications, high repetition rate operation will be necessary to generate and sustain the high average flux needed for many of the most exciting applications of laser-driven ion sources. Further, to navigate through the high-dimensional space of laser and target parameters towards experimental optima, it is essential to develop ion acceleration platforms compatible with machine learning  learning techniques and capable of autonomous real-time optimization. Here we present a multi-Hz ion acceleration platform employing a liquid sheet jet target. We characterize the laser-plasma interaction and the laser-driven proton beam across a variety of key parameters governing the interaction using an extensive suite of online diagnostics. We also demonstrate real-time, closed-loop optimization of the ion beam maximum energy by tuning the laser wavefront using a Bayesian optimization scheme. This approach increased the maximum proton energy by 11\% compared to a manually-optimized wavefront by enhancing the energy concentration within the laser focal spot, demonstrating the potential for closed-loop optimization schemes to tune future ion accelerators for robust high repetition rate operation.
\end{abstract}

\maketitle

\section{Introduction}
Beams of protons accelerated from interactions between short-pulse, high-intensity lasers and thin, solid-density targets have exhibited a variety of exciting properties, including high peak fluxes\cite{Snavely2000}, maximum proton energies exceeding \qty{100}{\mega\electronvolt}\cite{Higginson2018,Ziegler2024,Shou2025}, and low emittances surpassing those available using conventional accelerators\cite{Cowan2004}. Laser-driven proton accelerators are therefore promising candidates for a variety of future applications including tumor therapy\cite{Bulanov2002,Malka2004}, radioisotope generation\cite{Sun2021,Rodrigues2024}, inertial fusion energy\cite{Roth2001,Fernandez2009}, and the production of secondary particle beams including neutrons or alpha particles\cite{Bonvalet2021,Lancaster2004}.

Target normal sheath acceleration (TNSA) is the most widely studied mechanism of laser-driven proton acceleration from overdense targets and has been the subject of substantial theoretical and experimental interest over the last quarter-century\cite{Snavely2000, Wilks2001, Maksimchuk2000, Mora2003, Macchi2013}. In TNSA, a high-power, short-pulse laser focused to relativistic intensities ($\gtrapprox$\SI[per-mode=symbol]{1e18}{\watt\per\square\centi\meter}) irradiates a solid-density target with micron-scale thickness. Electrons liberated from the target front surface by the rising edge of the laser pulse are accelerated to relativistic energies and stream into the target. Part of the hot electron population escapes the target rear surface, establishing a quasi-static sheath electric field via charge separation. This field can have magnitudes up to \unit[per-mode=symbol]{\tera\volt\per\meter}, helping to confine the remaining hot electrons to the target bulk, where they recirculate, further heating the bulk material and increasing the strength of the sheath field. The rear surface layer of the target is almost immediately field-ionized and ions are accelerated normal to the target surface by the sheath field. Protons and light ions, primarily originating from the thin hydrocarbon layer adsorbed onto the target surface, are preferentially accelerated due to their large charge-to-mass ratio\cite{Wilks2001, Mora2003}. In continuously refreshing targets such as thin liquid sheets where no persistent hydrocarbon layer accumulates, light ions from the target rear surface will predominate due to their proximity to the accelerating sheath field. The influence of laser properties including energy, pulse duration, and temporal contrast\cite{Fourmaux2013,Flippo2008,Noaman-ul-Haq2018,Dover2020,McKenna2006,Ceccotti2007,Macchi2013} on TNSA proton spectra has been studied in detail, but typically over a limited parameter space in single-shot experiments conducted at repetition rates ${\ll}1\,\si{\Hz}$.

Fully realizing the promise of laser-driven proton accelerators will require operation at high repetition rate (HRR) from the \unit{\Hz} to the \unit{\kilo\Hz} level to robustly optimize their performance. HRR operation creates an application space for closed-loop optimization of laser-driven proton accelerators, enabling the use of active feedback and machine learning techniques to tune the beam properties as required within the high-dimensional, nonlinear space of laser and target parameters\cite{Ma2021,Hatfield2021}. These techniques may provide access to higher proton energies using existing laser and target hardware, or permit smaller laser systems to achieve proton energies currently only accessible with more energetic drivers\cite{Loughran2023}. Further, HRR operation with \unit{\peta\watt}-class laser drivers is needed for laser-driven proton sources to deliver the beam currents and energies (\unit{\micro\ampere} with maximum energies of tens of \unit{\mega\electronvolt}) required for demanding applications in isotope generation or the development of hybrid accelerators coupling laser-accelerated particle beams to RF accelerators\cite{Sun2021,Rodrigues2024,Krushelnick2000,Antici2008}.

Despite the recent proliferation of HRR PW-class laser systems worldwide\cite{Danson2019}, target development for HRR laser-driven proton acceleration remains an area of active research. These targets must overcome numerous practical challenges, including position stability within the Rayleigh range of the final focusing mirror (typically $\leq$\qty{10}{\micro\meter} for the small F-number optics used in these experiments\cite{Ohland2021}) and robustness against the harsh radiation environment generated by the laser-target interaction\cite{Prencipe2017}. In applications that require sustained operation (many hours to days) it is also essential to limit target debris, since buildup on focusing optics can cause damage and degrade laser beam quality, and to use easily replenishable target materials\cite{Prencipe2017}. A variety of potential targets that satisfy some or all of these constraints have been developed, including solid tape targets\cite{McKenna2002, Nayuki2003,Condamine2021,Xu2023,Zeraouli2023}, cryogenic liquid and gas jet targets\cite{Kim2016, Margarone2016,Curry2020a,Chagovets2022}, and liquid sheet jet targets\cite{Morrison2018,George2019,Puyuelo-Valdes2022,Crissman2022,Treffert2022,Treffert2022a,Cao2023,Fule2024,Peng2024,He2025}.

Liquid sheet jet targets present a particularly promising target platform due to their ability to reach smaller thicknesses than tape while retaining solid densities and producing minimal debris. These targets thus satisfy all the requirements for sustained HRR operation using petawatt-class lasers, allowing the exploration of new ion acceleration regimes and the demonstration of closed-loop ion beam optimization. We recently demonstrated that water sheet jet targets could be used to generate very low-divergence ion beams through self-collimation of the laser-driven proton beam in background water vapor produced by the target, relaxing the constraints on beam focusing hardware needed to obtain high-flux laser-driven proton beams at HRR\cite{Streeter2025}. These interactions delivered an average flux of $(8.9\pm2.0)\times10^{9}$ protons per \unit{\joule} of laser energy over 50 consecutive shots, with ${\sim}10^{11}$ protons/\unit{\steradian} with energies between 1 and \qty{2}{\mega\electronvolt}.

In this work, we report on a variety of additional measurements characterizing the ion acceleration platform used in Ref.~\citenum{Streeter2025} and demonstrating its suitability for sustained HRR operation. We present optical shadowgraphy of the laser-plasma interaction at delays up to ${\sim}700$~\unit{\pico\second}, monitoring the growth of plasma instabilities and the development of a laser-driven shock. We also show that p-polarized laser pulses, compared to s-polarized or circularly-polarized laser pulses, improve the proton maximum energy by a factor of three and the peak dose by an order of magnitude, achieving peak single-shot doses of up to \qty{30}{\gray}. Finally, we demonstrate real-time, closed loop optimization of the laser-driven proton source using a Bayesian optimization scheme operating at a repetition rate of $\qty{1}{\Hz}$. The optimizer adjusted the laser wavefront to improve the energy containment within the focal spot, increasing the proton maximum energy by 11\%. These results pave the way towards application-ready proton sources delivering multi-\unit{\mega\electronvolt} peak energies with high single-shot doses at multi-\unit{\Hz} repetition rates, actively stabilized and optimized using machine learning.

\section{Experimental Results}
\subsection{Setup}
\begin{figure}
    \centering
    \includegraphics[width=\linewidth]{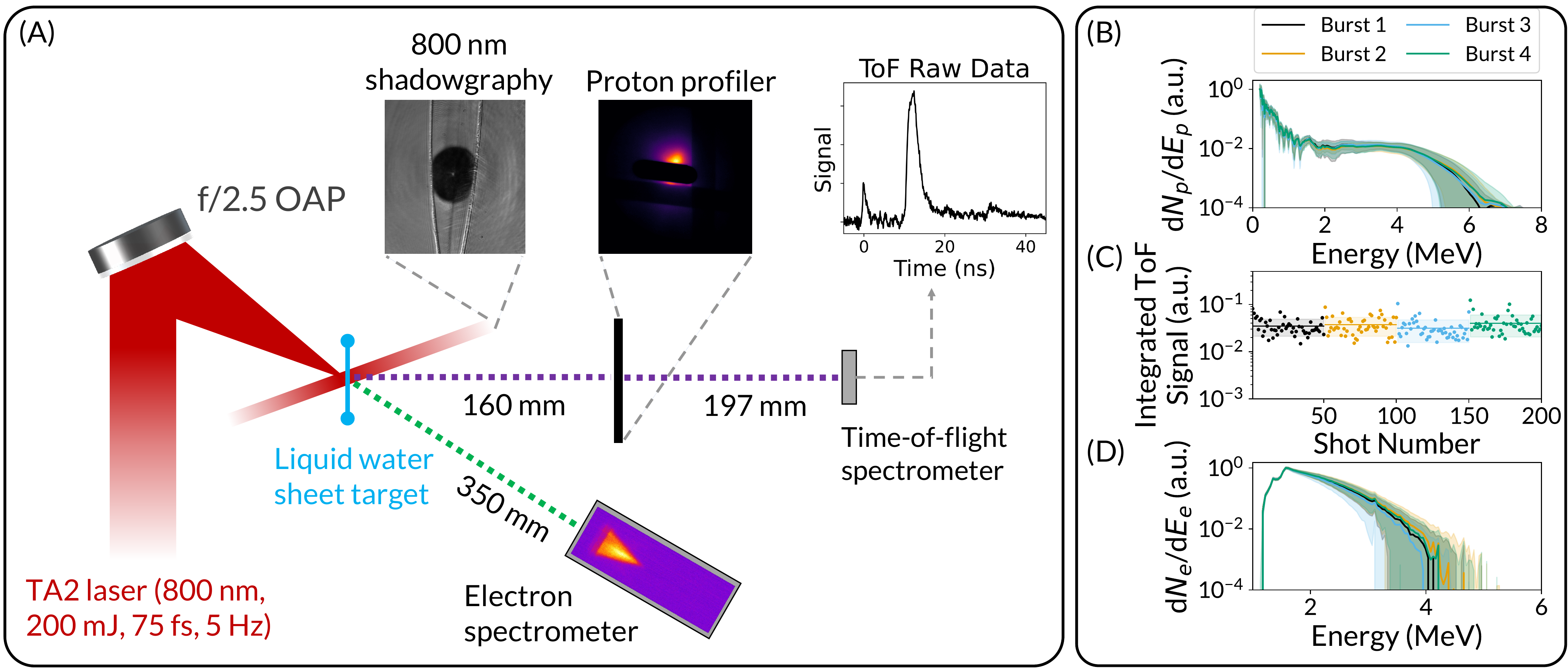}
    \caption{(A): An illustration of the experimental setup showing the high-power laser, the liquid sheet target, and the diagnostics used to characterize the laser-plasma interaction and the energetic electrons and protons generated in the interaction. The main laser irradiated a ($600\pm100$)~\unit{\nano\meter} thick liquid water sheet jet at a repetition rate of up to \qty{5}{\Hz}. Electrons and protons accelerated in the interaction were characterized using a magnetic electron spectrometer, a scintillator-based proton beam profiler, and a diamond diode time-of-flight (ToF) spectrometer. An independently timed \unit{\femto\second}-duration laser pulse collected shadowgraphic images of the interaction. (B) Burst-averaged proton spectra from the ToF diode collected at \qty{1}{\Hz} in four consecutive bursts composed of 50 shots each. (C) Integrated proton flux measured by the ToF diode on each shot in the four bursts. The solid horizontal lines denote the burst averages, with the shaded regions indicating $\pm1$ standard deviation. (D) Burst-averaged electron spectra collected at \qty{1}{\Hz} in the same four bursts shown in (B) and (C).}
    \label{fig:setupfig}
\end{figure}

Fig.~\ref{fig:setupfig} schematically illustrates the configuration of the laser, target, and experimental diagnostics. The high-power laser delivered pulses containing up to \qty{200}{\milli\joule} in a typical FWHM duration of $75\pm5$~\unit{\femto\second}, which were focused to an intensity of up to $3\times10^{19}$~\qty[per-mode=symbol]{}{\watt\per\square\centi\meter} onto the surface of a continuously flowing liquid H\textsubscript{2}O sheet jet\cite{Treffert2022a} at a repetition rate of up to \qty{5}{\Hz}. Due to limitations of the data acquisition system, much of the data presented here was acquired at \qty{1}{\Hz}. The laser was incident at 30$^\circ$ with respect to the target normal. The target thickness was measured using white light interferometry to be $600\pm100$~\unit{\nano\meter} at the interaction point \qty{2.8}{\milli\meter} below the bottom of the nozzle, where the error bar denotes measurement uncertainty rather than thickness variation.

Relativistic ``hot'' electrons accelerated in high-intensity laser-target interactions are primarily responsible for transferring the laser energy to the energetic proton beam. As such, the characteristics of these electrons are strongly correlated with the properties of the proton beam. Here, the energy distribution of the electrons escaping the target along the laser forward direction was sampled using a permanent magnet-based electron spectrometer. The spectrometer was equipped with a Lanex screen and a charge-coupled device (CCD) camera for HRR operation. 

The protons escaping the target, which reached maximum energies of several \unit{\mega\electronvolt}, were characterized both spatially and spectrally. The spatial distribution of the low-energy component of the proton beam was measured using a scintillator-based proton beam profiler centered on the target rear-surface normal and placed \qty{160}{\milli\meter} from the interaction point. The majority of the proton profiler surface was covered by a \qty{12}{\micro\meter} thick aluminized Mylar shield which blocked light from the laser and signal from heavy ions, meaning that the scintillator was most sensitive to protons with energies of \qty{1.1}{\mega\electronvolt}. The camera signal from the proton beam profiler was absolutely calibrated into a proton dose using radiochromic film. The scintillator had a slot which provided a line of sight to a diamond detector time-of-flight (ToF) spectrometer placed \qty{357}{\milli\meter} from the interaction that was used to determine the energy distribution of the laser-accelerated protons. Additionally, an independent \qty{800}{\nano\meter} laser pulse (pulse energy ${\sim}1$~\unit{\milli\joule}, pulse duration ${\sim}40$~\unit{\femto\second}) with variable delay was used as a shadowgraphic probe of the laser-plasma interaction. See the \hyperref[sec:methods]{Methods} section for additional details regarding the experimental setup.

We observed that the liquid sheet target facilitated high-intensity laser-target interactions that were stable shot-to-shot, leading to reproducible hot electron and proton signals. In Fig.~\ref{fig:setupfig}(B--D), we show proton spectra, integrated proton fluxes, and electron spectra obtained at a repetition rate of \qty{1}{\Hz} in four consecutive ``bursts'' composed of 50 shots each. The relative standard deviation of the integrated proton flux (standard deviation divided by the mean) for each of the bursts is at most 50\%, indicating that typical fluctuations in the total proton flux produced in the interaction along the line of sight of the ToF spectrometer are no more than a factor of two. Similarly, we observe that the proton and electron spectra are very consistent across all four bursts, with slightly more energetic electrons in burst 2 but not beyond the $1\sigma$ contours from the other bursts. The shot-to-shot stability of this laser-driven ion acceleration platform makes it well-suited for collecting large datasets composed of scans across a variety of experimental parameters. 

\subsection{Target Platform Characterization for HRR operation}
\begin{figure}
    \centering
    \includegraphics[scale=0.5]{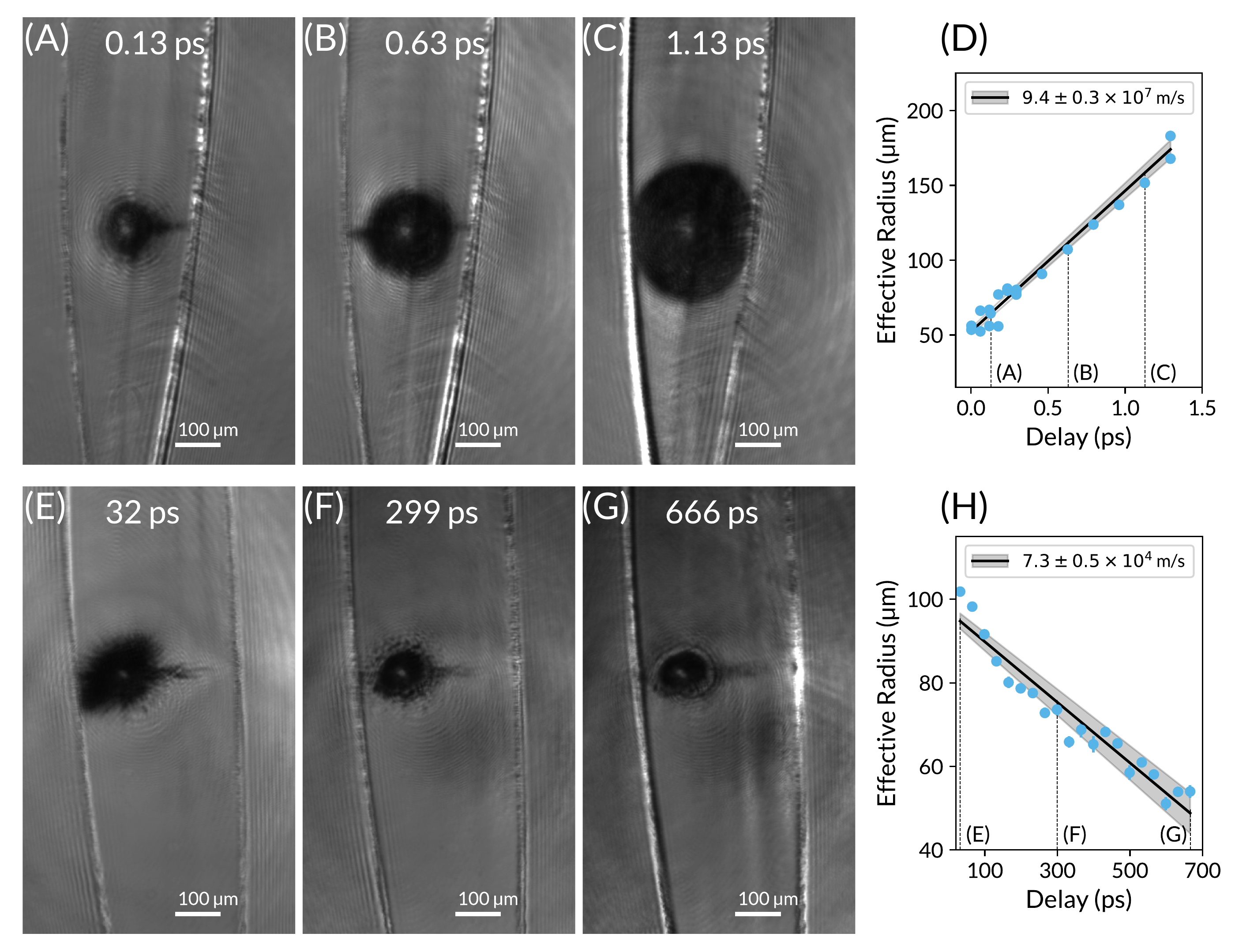}
    \caption{(A)--(C): Shadowgraphic images from the \qty{800}{\nano\meter} probe diagnostic showing plasma expansion within approximately the first \unit{\pico\second} after the laser-plasma interaction. Plasma electron densities above the critical density are visible as an opaque region in the images. Images and the extracted data are averaged across approximately 15 shots. (D): The effective radius (geometric mean of the semimajor and semiminor axes of a fitted ellipse) of the laser-produced overdense plasma plotted as a function of time at early times after the laser-plasma interaction. Error bars (typically smaller than the marker size) indicate $\pm$~1 standard error of the mean. A linear best fit to the data indicates that the overcritical plasma region expanded at a rate of approximately $9.4\times10^{7}$~\unit[per-mode=symbol]{\meter\per\second}. (E--G): Shadowgraphic images showing the late-time behavior of the interaction region. The plasma has shrunk considerably due to recombination and expansion into the vacuum, and at the latest times a ``rim'' feature corresponding to an outgoing shock wave can be observed. (H): The effective radius of the laser-produced overdense plasma as a function of time plotted over several hundred \unit{\pico\second} after the interaction. A linear best fit to the data indicates that the overcritical plasma shrunk at a rate of approximately $7.3\times10^{4}$~\unit[per-mode=symbol]{\meter\per\second} during this time period.}
    \label{fig:probefig}
\end{figure}

The use of an independent probe beam also enabled detailed monitoring of the plasma dynamics after the high-intensity laser-matter interaction. Optical shadowgraphy of the early-time dynamics can provide insights into the target microphysics, while probing at long delay times can help determine the time required for the target to recover from the laser shot and hence the maximum repetition rate achievable by the target. Here, we used the low-energy, short-pulse probe laser beam described above to backlight the region of interest, then varied the relative delay between the high-intensity interaction and the arrival of the probe pulse to monitor the plasma expansion and observe the formation of a laser-generated shock. Continuous operation at \qty{1}{\Hz} allowed us to obtain high-resolution temporal scans with repeated acquisitions at each point in parameter space, improving the statistical power of the results. These results are presented in Fig.~\ref{fig:probefig}, which shows a series of images illustrating both the early-time expansion of the hot plasma from the laser-target interaction and the late-time emergence of a shock front expanding away from the interaction. The plasma is visible as a darkened region in these images. We expect that the observed opacity, particularly at the earliest times, is dominated by reflection and absorption of the probe beam from overdense plasma with an electron density of at least $n_{c}^{800\text{ nm}}=1.74\times10^{21}$ \unit{\per\centi\meter\cubed}, since ionization of the solid-density target occurs quickly. Subsequently, refractive effects may contribute near the transverse edges of the expanding plasma.

 In Fig.~\ref{fig:probefig}(A)-(C), the expansion of the plasma region away from the interaction point within the first \unit{\pico\second} after the high-intensity interaction is visible. A series of measurements of the spatial extent of the opaque plasma allows us to calculate the expansion rate of the ionized region to be approximately $9.4\times10^{7}$~\unit[per-mode=symbol]{\meter\per\second} (${\sim}0.3c$). This rapid expansion is likely driven by collisional ionization from hot electrons generated in the initial laser-plasma interaction. Previous work has estimated that the hot electrons expand in the target plane at roughly $0.5c$ as a result of collisional effects and instabilities\cite{Ruyer2020}, in fair agreement with our experimental observation. The overall scale of the ionized region is also consistent with other measurements obtained using significantly lower-energy laser pulses, which found that the overdense plasma region expanded to a diameter of approximately \qty{100}{\micro\meter} in less than \qty{5}{\pico\second}\cite{Morrison2018}.

By \qty{30}{\pico\second} after the interaction, the initial plasma expansion has recombined or rarefied and filamentary structures are visible at the interface between the hot plasma and the cold target bulk (Fig.~\ref{fig:probefig}(E)). In a different experiment using shadowgraphic probing of relativistically intense laser-plasma interactions, similar features were observed after 10 to \qty{100}{\pico\second}. These features were attributed to the formation of a radial Weibel-type instability as the hot electrons expanding from the interaction counter-streamed against the cold return current\cite{Ngirmang2020}. Another experiment, which instead used proton radiography to probe magnetic field structures in the target plane, also found evidence for the development of filamentation instabilities that could extend over tens of \unit{\pico\second} and hundreds of \unit{\micro\meter}\cite{Ruyer2020}. The authors of Ref.~\citenum{Ruyer2020} found that the instability growth rate was of order \unit{\per\pico\second}, with filament wavelengths $\lambda_{p}$ of order \qty{100}{\micro\meter}; we observe $\lambda_{p}\approx20$ \unit{\micro\meter} (Fig.~\ref{fig:probefig}(E--F)). According to the model proposed in Ref.~\citenum{Ruyer2020}, this difference could result from differing degrees of electron momentum anisotropy between the two experiments, perhaps due to the substantial differences in the laser and target conditions used. Further study of the filamentary features observed here to clarify the plasma instability responsible for their emergence and growth is of fundamental interest for understanding high intensity laser-plasma interactions, but will require a significant numerical modeling campaign beyond the scope of this work. 

By approximately \qty{400}{\pico\second}, the filamentary structures have collapsed into a growing shock front, and at the latest times probed the radius of the region perturbed by the high intensity laser-target interaction has begun to saturate at approximately \qty{50}{\micro\meter}  (Fig.~\ref{fig:probefig}(G--H)). A recent XFEL study performing direct x-ray imaging of short pulse laser-heated wires also observed a similar collapse of plasma filaments visible on \unit{\pico\second} timescales into a shock front\cite{Schoenwaelder2025}. No bulk motion of the perturbed region is visible on the timescales shown here; given the jet flow velocity of approximately \qty[per-mode=symbol]{10}{\meter\per\second}, the surface has moved only a few tens of \unit{\nano\meter}. Further growth of the perturbed region occurred from expansion of the shock front through the target bulk on longer timescales inaccessible to probing during this experiment. Based on shadowgraphic probing of laser-irradiated liquid jet targets at longer delay times reported elsewhere\cite{George2019}, we anticipate the shock to expand and perturb a region of up to several hundred microns in diameter on the \unit{\micro\second} timescale. While this comparison suggests that operation at the laser parameters presented here may be scalable to repetition rates of order \qty{10}{\kilo\Hz}, future studies investigating the late-time fluid dynamics of the irradiated target will require alternative probe beams capable of \unit{\micro\second} delays or the use of high-speed photography. These studies are of significant interest in demonstrating the viability of plasma-based accelerators\cite{DArcy2022} and will be necessary for scaling laser-driven ion sources to \unit{\kilo\Hz} repetition rates.

\subsection{1D and higher-dimensional grid scans}
In addition to enabling detailed characterization of the target evolution at fixed laser irradiation conditions, the use of a HRR platform allowed us to perform scans over various experimental parameters as well as to quantify the stability of performance across the scan by acquiring multiple shots at each point in parameter space.

\begin{figure}
    \centering
    \includegraphics[scale=0.45]{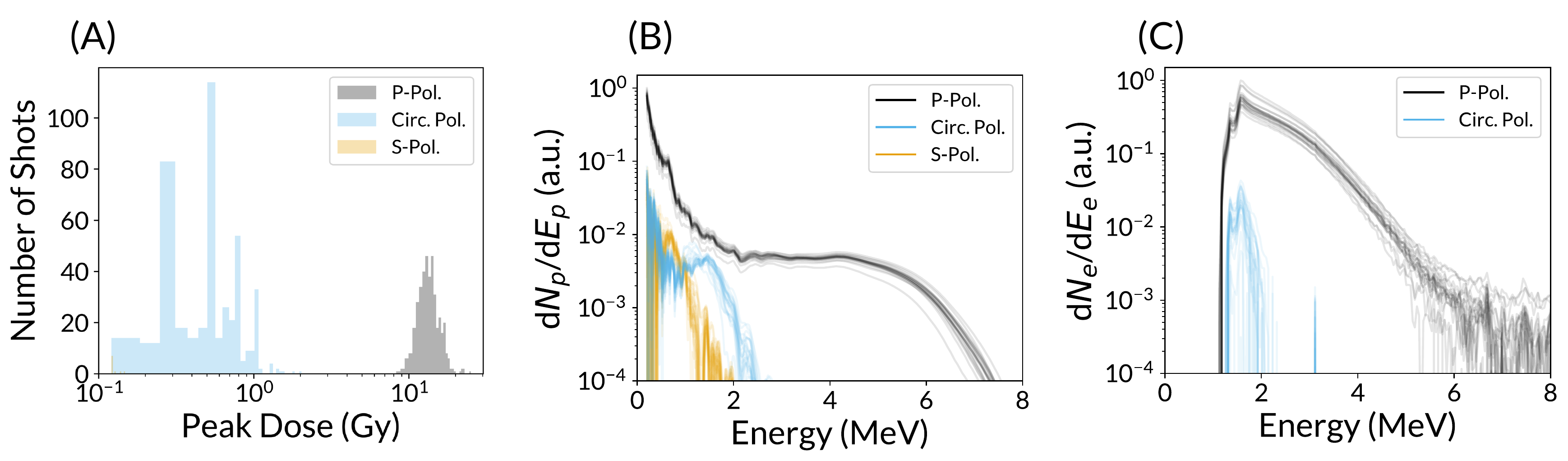}
    \caption{(A) Histogram of the peak proton dose deposited in the scintillator in a single shot for proton beams accelerated using p-, s-, and circularly-polarized laser pulses. Only a few s-polarized interactions generated enough signal on the scintillator to be observable over the background. (B) Proton energy spectra measured by the ToF spectrometer for proton beams accelerated using p-, s-, and circularly-polarized laser pulses. (C) Electron energy spectra obtained when using p-polarized and circularly polarized laser pulses. S-polarized pulses accelerated too few electrons for signal to be observed on the spectrometer. Each trace shown in (B), (C) is the average of one burst of 20 shots.}
    \label{fig:polarization}
\end{figure}

The polarization of the high-intensity laser is a key parameter governing the laser-target coupling. P-polarization, in which the electric field component of the incident beam oscillates in the plane of the target, has been shown to enhance laser-target coupling through a variety of mechanisms in both theoretical and experimental studies\cite{Wilks1997,Cerchez2008,Adak2021}. In Fig.~\ref{fig:polarization} we present measurements of the proton dose, proton energy spectra, and electron energy spectra obtained in relativistically intense interactions using s-polarized, p-polarized, and circularly polarized laser pulses. The impact of changing the polarization was significant, particularly in the particle flux: the proton dose generated from p-polarized laser pulses was typically at least an order of magnitude greater than that generated using circularly polarized pulses, and nearly two orders of magnitude greater than that generated using s-polarized laser pulses, which typically fell below the noise floor of the detector.

P-polarized pulses also consistently yielded maximum proton energies approximately $3\times$ higher than those produced using either s- or circularly-polarized laser pulses. S-polarized pulses did not accelerate a sufficiently large population of electrons to energies greater than the \qty{1}{\mega\electronvolt} lower bound of the electron spectrometer for signal to be visible on the diagnostic. Since the experiment was performed in a regime of laser intensities and target densities in which TNSA is expected to be the dominant proton acceleration mechanism\cite{Macchi2013}, this observation is consistent with reduced proton yields due to the need for hot electrons to mediate the transfer of laser energy to accelerated protons.

Our observations of the effect of the laser polarization are also consistent with previous theoretical and experimental work. It has long been recognized that, for relativistically intense laser pulses incident on a sharp plasma density gradient, p-polarization of the pulse allows the laser field and ion restoring force from the target bulk to push and pull electrons back and forth across the gradient\cite{Brunel1987,Wilks1997}. In this heating mechanism, known as vacuum heating, the electron oscillation amplitude grows with time, increasing the laser energy coupled into the electrons and thus the electron heating. Vacuum heating requires high-contrast laser pulses in order to mitigate pre-expansion of the target, which would otherwise decrease the relative coupling efficiency of p-polarized laser pulses into the target\cite{Wilks1997}. The threefold improvement in proton energy we observed for p-polarized pulses compared to s-polarized pulses is less than that observed in Ref.~\citenum{Ceccotti2007}, which we may explain by the lower contrast of the laser used in this study compared to their case. Another recent work using similar laser conditions to those presented here\cite{Almassarani2021} also observed improved proton energies from p-polarized laser pulses compared to s-polarized pulses, but only by a factor of roughly 1.5.

For laser pulses normally incident on the target, circular polarization is expected to significantly suppress electron heating by contributions from the $\mathbf{j}\times\mathbf{B
}$ force\cite{Wilks1997}. As a result of the oblique incidence angle used in this study, however, the circularly polarized pulses had a p-polarized component with an electric field amplitude $\sqrt{2}$ times lower than in the purely p-polarized case. Consistent with our observed electron and proton spectra, these pulses yielded reduced hot electron generation and lower proton peak energies and fluxes than purely p-polarized pulses, but not to the same degree as s-polarized pulses.

\begin{figure}
    \centering
    \includegraphics[width=\linewidth]{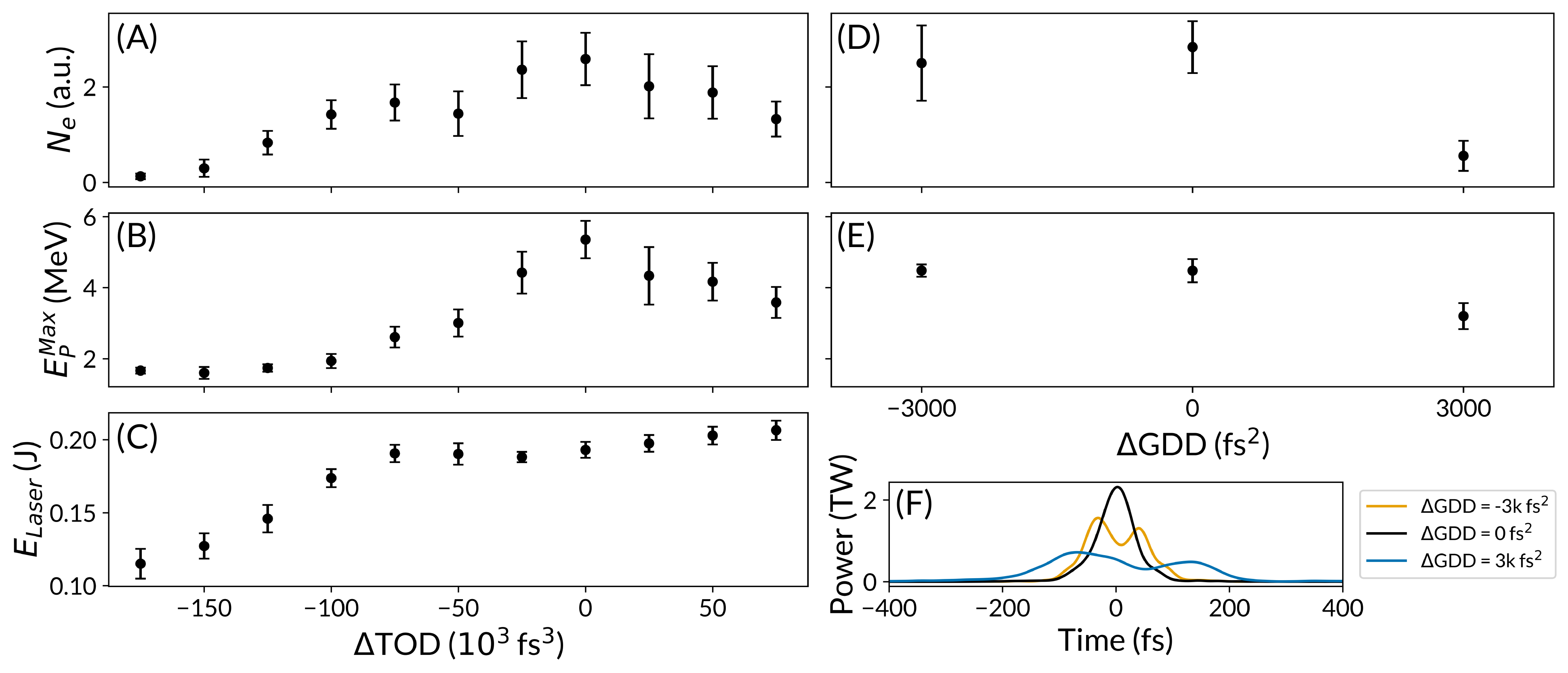}
    \caption{(A) Total electron signal measured by the electron spectrometer as a function of the change in the third-order dispersion (TOD) applied to the laser pulse, where $\Delta TOD=0$ is defined as the TOD required to obtain the best-compressed pulse duration. (B) Proton maximum energy as measured by the ToF spectrometer as a function of the change in TOD applied to the laser pulse. (C) Laser pulse energy as a function of the change in TOD. Note that large changes in the TOD substantially reduced the laser pulse energy. (D) Total electron signal as a function of the change in the group delay dispersion (GDD) applied to the laser pulse. (E) Proton maximum energy as a function of the change in GDD applied to the laser pulse. (F) Temporal profiles of the pulses corresponding to $\Delta GDD=0$ (best compression), +\qty{3000}{\femto\second\squared}, and -\qty{3000}{\femto\second\squared}. Each point shown is the average of one burst of 10 shots. Error bars represent $\pm$ one standard deviation.}
    \label{fig:fig:dazzler}
\end{figure}

Shaping the laser pulse on femtosecond timescales by modifying spectral phase components of the pulse has also been observed to significantly impact laser-driven ion acceleration performance\cite{Ziegler2021,Psikal2024}. In our experiment, an acousto-optic programmable filter (AOPDF, Fastlite Dazzler) was capable of adjusting the group delay dispersion (GDD), third-order dispersion (TOD), and fourth-order dispersion (FOD) applied to each laser pulse. Figure~\ref{fig:fig:dazzler} presents our observations of the impact of adding either GDD or TOD to the laser pulse on the total number of fast electrons observed by the electron spectrometer and the proton maximum energy measured by the ToF spectrometer. In general, the best electron yields and the highest proton energies were obtained with $\Delta\textrm{GDD}=0$ and $\Delta\text{TOD}=0$, which was the nominal best-compression configuration. Positive values of $\Delta\text{TOD}$, which have been observed to improve ion acceleration performance by extending the rising edge of the laser pulse\cite{Ziegler2021,Psikal2024}, did not improve either the number of electrons accelerated or the maximum proton energy. In previous work joint variation of GDD and TOD typically yielded the greatest improvements in ion acceleration performance, but these parameters were only tuned individually in this study. We also observed substantial decreases in the laser pulse energy for large negative values of $\Delta\text{TOD}$ due to the accompanying spectral distortion of the pulse, complicating comparisons to values of $\Delta\text{TOD}$ below -50,000~\unit{\femto\second\cubed}. We observed smaller effects from changing the pulse GDD than reported in Ref.~\citenum{Ziegler2021}, though in other work using submicron foil targets the ion energy was robust against or even benefited from significant changes in the pulse duration caused by chirping the pulse\cite{Permogorov2022,Tayyab2018}.

\begin{figure}
    \centering
    \includegraphics[width=\linewidth]{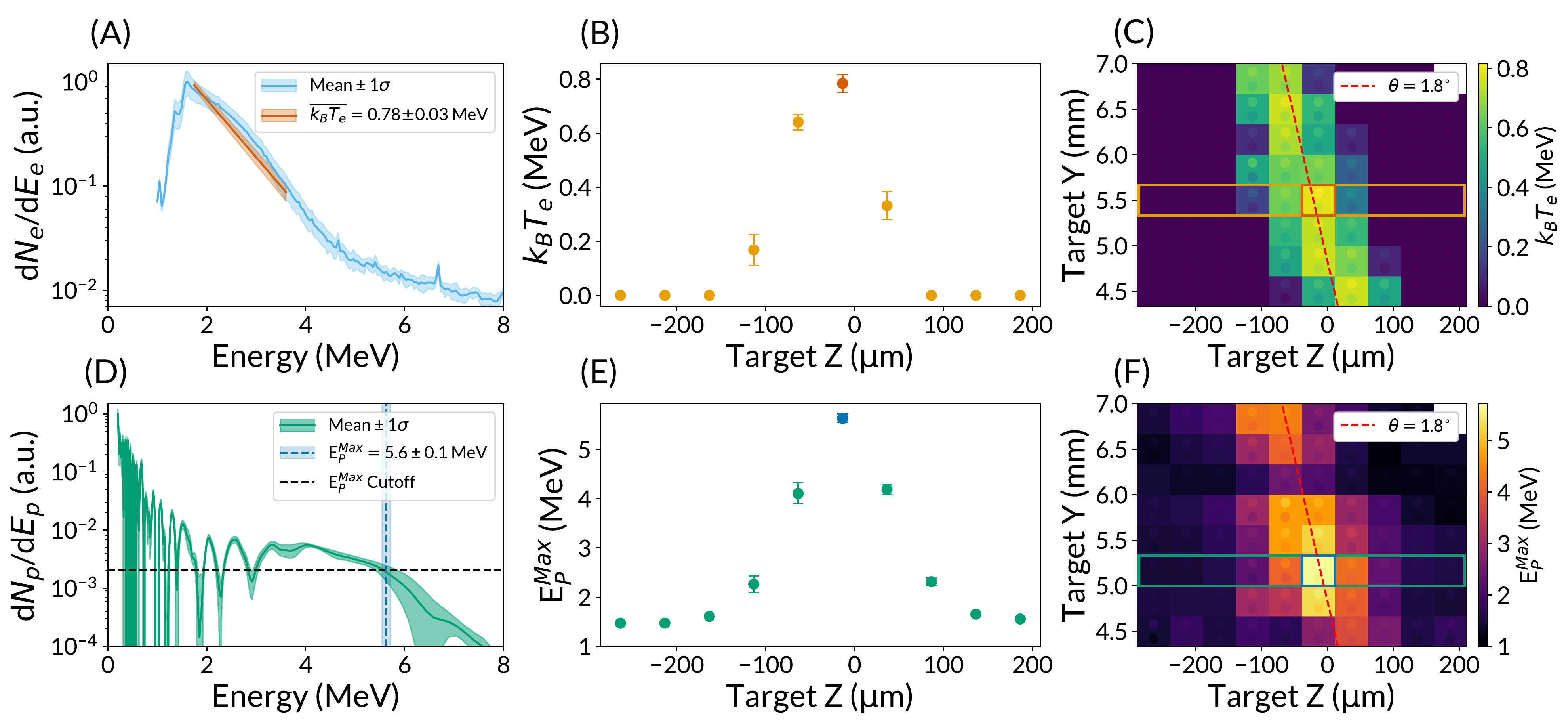}
    \caption{(A) A series of single-shot electron spectra collected at a fixed set of conditions, providing an estimation of the measurement uncertainty. The exponential tail of each spectrum is fitted to an equation of the form $\text{exp}\left(-E/k_{B}T_{e}\right)$ to obtain the effective temperature of the distribution $k_{B}T_{e}$. (B) The mean electron temperatures measured as shown in (A) from a 1D scan across the target position along the laser axis (Z). (C) The mean electron temperatures obtained from a 2D scan of Z and target height (Y). Each square on the grid represents one point in (Z, Y) parameter space. Inset colored circles indicate the mean value plus (upper) or minus (lower) one standard error of the mean. Overlaid colored boxes indicate the 1D subspace and single point in parameter space corresponding to (B) and (A) respectively. (D) A series of single-shot proton spectra obtained from the ToF spectrometer at a fixed set of conditions. The maximum proton energy $E_{P}^{Max}$ is determined by calculating the highest energy at which the signal crosses a specified cutoff value. (E) Maximum proton energies obtained from a 1D scan along Z. (F) Maximum proton energies measured in the same 2D scan of Z and Y shown in (C). Overlaid colored boxes indicate the 1D subspace and single point in parameter space corresponding to (E) and (D) respectively. Data points are averaged across 10 shots per burst, and error bars represent $\pm$ one standard error of the mean.}
    \label{fig:dimensionality}
\end{figure}

For Hz-scale operation it is also possible to perform high-resolution or multidimensional parameter scans in a reasonable timescale while still obtaining good statistics at each point in parameter space. These scans might vary laser parameters such as energy or pulse shape, target parameters such as thickness, or combinations thereof. In Fig.~\ref{fig:dimensionality}, we show a series of data obtained from the electron spectrometer and proton ToF spectrometer with a ``zero-dimensional'' scan (fixed position in parameter space), as well as one- and two-dimensional parameter scans over the target position. Data acquired at a fixed point in parameter space make it possible to quantify the uncertainty in the experimental measurement as a result of laser or target fluctuations. Due to the highly nonlinear physics of intense laser-plasma interactions, small shot-to-shot variations in the interaction conditions can lead to large changes in the resulting plasma properties and consequently the ion beam parameters, making uncertainty quantification from repeated shots at fixed conditions crucial for reliable measurements.

Scans over a single parameter, as typically performed in single-shot experiments, can be used to determine an empirical experimental optimum of some parameter of interest (e.g., proton or electron flux, maximum energy, or temperature as in Figs.~\ref{fig:polarization} and \ref{fig:fig:dazzler}). By scanning over two parameters, it is possible to identify coupling between them---such as the target position along the laser axis (Z) and height (Y). In Fig.~\ref{fig:dimensionality}, the observed correlation between high-signal points in (Z, Y) space visible in both the electron temperature (Fig.~\ref{fig:dimensionality}(C)) and the maximum proton energy (Fig.~\ref{fig:dimensionality}(F)) can be attributed to a tilt of the target in the vertical direction of approximately $1.8^\circ$. This observation was generally consistent with post-experiment analysis of the proton beam pointing on the proton profiling diagnostic, demonstrating the value of a comprehensive ion diagnostic suite collecting both spatial and spectral information to confirm measurement consistency. During experimental operation, such observations provide opportunities to optimize target operation (such as by correcting the tilt observed here) in real time, or to move to parameter space optima that would have been missed in single 1D parameter scans or sequential 1D scans of two parameters\cite{Shalloo2020}. It is important to emphasize that these scans require shot numbers that are infeasible without HRR operation. For example, the 2D grid scans shown in Fig.~\ref{fig:dimensionality}(C) and (F) are composed of 79 ``bursts,'' each containing 10 shots. At a repetition rate of \qty{1}{\Hz}, the scan could be completed in roughly 30 minutes of experimental time including parameter changes between bursts, underscoring an additional advantage of HRR operation: while large datasets can be accumulated at lower shot rates given a longer time window, temporal evolution of the laser system (e.g. thermal drifts or alignment fluctuations) may begin to influence the dataset over these long durations. The ability to quickly perform 1D automated scans can also help minimize the impact of systematic fluctuations in the laser system by enabling quick scans of key variables (such as target focal position or laser pulse shape) on a regular basis during data acquisition for comparison to previous benchmarks.

\subsection{Real-Time Optimization}
Even with HRR operation, fine grid scans over more than two or at most three parameters quickly become experimentally infeasible: the timescale to complete the scan grows exponentially with the number of parameters. Strong dependence of the laser-driven proton beam quality on the scan parameters also means that most of the data gathered in a multi-dimensional grid scan would be in regions of parameter space with poor proton acceleration performance, far from experimental optima. Finally, coupling between the parameters could mean that regions of parameter space are redundantly sampled, providing no additional insights into optimization of the proton beam. So, in settings where it is necessary to efficiently optimize laser-driven proton beams over the many control parameters of both the drive laser and the target, machine learning-based techniques for high-dimensional optimization are vital. Here, we employed Bayesian optimization (BO) to perform real-time, in-situ optimization of the laser-driven proton beam, incorporating Gaussian process regression to form a surrogate model of the parameter space with quantified uncertainty.

BO is a widely-used technique for optimizing multivariate functions which are expensive to evaluate and whose form is generally not well-constrained\cite{Shahriari2016,Frazier2018,Dopp2023}. In BO, one begins with a prior model of the objective function, which is then updated with successive observations of experimental data to obtain a posterior model which more accurately matches the observations. After each observation, the model can be consulted to obtain an updated prediction for the most likely position of the optimum within the parameter space. Objective functions in BO are commonly modeled using Gaussian processes; these models straightforwardly incorporate uncertainty estimates, making them a natural choice for experimental applications. BO has been used in both experiments and simulations of laser-driven particle acceleration to maximize the resulting energies or tune the acceleration process for other desired parameters\cite{Shalloo2020,Jalas2021,Dolier2022,Loughran2023,Irshad2024,Mariscal2024}.

\begin{figure}
    \centering
    \includegraphics[width=0.92\linewidth]{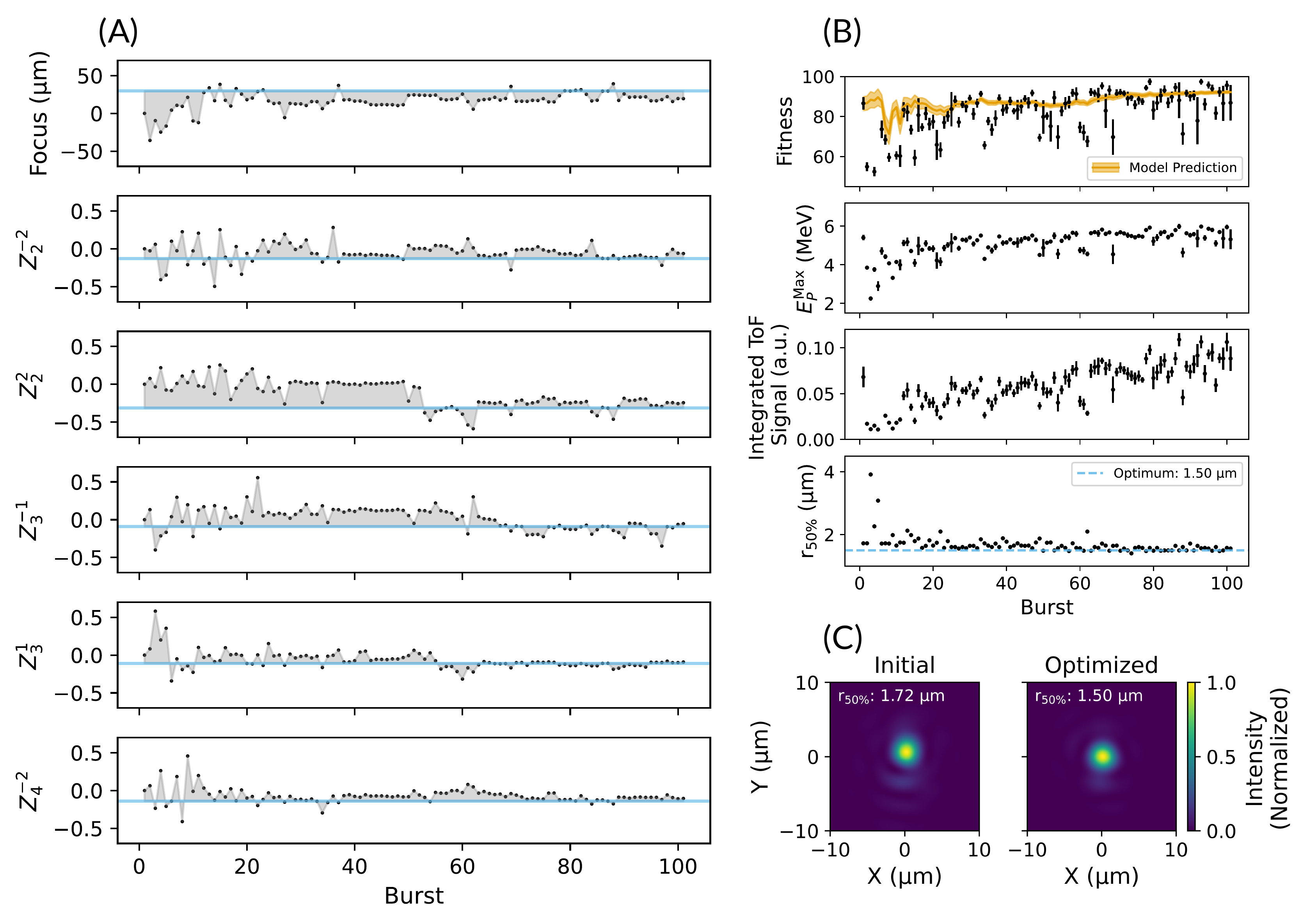}
    \caption{(A) Variation in the Zernike mode coefficients in the laser system deformable mirror over the course of the optimization. Each panel shows the values of a single parameter as black points. The parameter values corresponding to the global maximum ToF spectrometer energy are indicated with a horizontal blue line. The Zernike coefficients tuned were $Z^{0}_{2}$ (focus), $Z^{-2}_{2}$ (oblique astigmatism), $Z^{2}_{2}$ (vertical astigmatism), $Z^{-1}_{3}$ (vertical coma), $Z^{1}_{3}$ (horizontal coma), and $Z^{-2}_{4}$ (oblique second-order astigmatism). (B) Top: fitness values (black points) calculated by the optimizer after each burst of shots during the experiment (see text). Prior to each burst, the model predicted the fitness based on all data acquired up to that point (yellow line). Second from top: the maximum proton energy calculated from the ToF spectrometer for each burst, incorporating post-experiment recalibration. Third from top: the integrated proton flux observed on the calibrated ToF spectrometer for each burst. Bottom: the 50\% enclosed radius of the laser focal spot reconstructed from the Zernike coefficients applied using the deformable mirror for each burst of the optimization. Error bars indicate $\pm$ 1 standard error of the mean. (C) Reconstructed far-field intensity distributions of the focused laser, showing that the BO-optimized focal spot improved the energy containment compared to the initial manually-optimized focal spot.}
    \label{fig:optimization}
\end{figure}

In this experiment we used the BO platform described in Ref.~\citenum{Loughran2023}, expanded from that used in Ref.~\citenum{Shalloo2020}, to optimize a variety of different properties of the proton and hot electron beam including beam flux and maximum energy.  The platform could control subsets of the laser properties (e.g. energy, temporal pulse shape, wavefront, and polarization) and the target position. Fig.~\ref{fig:optimization} illustrates the use of BO to optimize the maximum energy of protons accelerated from the liquid sheet target as measured by the ToF spectrometer. In this dataset, the optimizer controlled the deformable mirror (DFM) and could adjust six Zernike mode coefficients impacting the laser wavefront: $Z^{0}_{2}$ (focus), $Z_{2}^{-2}$ (oblique astigmatism), $Z_{2}^{2}$ (vertical astigmatism), $Z_{3}^{-1}$ (vertical coma), $Z_{3}^{1}$ (horizontal coma), and $Z_{4}^{-2}$ (oblique second-order astigmatism). The effect of these parameters on laser-driven proton beam quality is complex, modulated by both the resulting laser peak intensity and the energy concentration within the focal spot. Combined with their suitability for HRR electronic control, these parameters were well-suited to autonomous optimization using BO.

We began the optimization with a manually-optimized best focus. During the optimization, the control system collected ten shots at \qty{1}{\Hz} (comprising one burst) and collated the data. The optimizer extracted the maximum proton energy from the ToF spectrometer for each shot (Fig.~\ref{fig:dimensionality}(D)), then scaled this value to obtain the ``fitness.'' The mean and standard error of the fitness values obtained in the burst were computed, then provided to the optimizer to obtain a new posterior model via Gaussian process regression. The updated posterior model was used to predict the optimal parameter values (i.e., the next set of values of the six Zernike coefficients) and the corresponding fitness by maximizing an expected improvement acquisition function. The optimizer could then instruct the control system to adjust the DFM appropriately and begin taking shots for the next burst. In the scan presented here, this process was repeated for 101 bursts before it was manually terminated, as no automated termination criteria were set.

Fig.~\ref{fig:optimization}(A) shows the values of these parameters as they were varied throughout the optimization. Each panel describes a single parameter, with the parameter values for each burst indicated with black points and the parameter values corresponding to the maximum ToF spectrometer energy obtained in the optimization indicated with a horizontal blue line. The initial model of the objective function was constructed from a random sampling of the parameter space, with the prior objective initially set to zero across the parameter space. In the first bursts after the start of the optimization, the optimizer explored an unfavorable position in parameter space, leading to a significant degradation in the energy containment within the focal spot (Fig.~\ref{fig:optimization}(B), bottom panel). At this position, the fitness decreased considerably, (Fig.~\ref{fig:optimization}(B), top panel), as did the maximum proton energy and flux (Fig.~\ref{fig:optimization}(B), middle two panels). A key advantage of BO is that all the acquired data is incorporated into the posterior model of the objective function, which provides the model of parameter space. So, even acquisitions of data in poorly performing regions of parameter space are of value to the model construction. After this initial venture into a poorly-performing region of parameter space the model returned to more favorable regions, and the remainder of the data acquisition was concentrated in regions of parameter space that generated proton beam maximum energies relatively near the final optimum.

Several key experimental observables are plotted over the course of the optimization in Fig.~\ref{fig:optimization}(B), showing that the proton beam maximum energy was generally fairly close to the initial value of \qty{5.4}{\mega\electronvolt}. The highest value obtained in the optimization was $6.0\pm0.1$~\unit{\mega\electronvolt}, 11\% higher than the initial maximum energy. This optimum was associated with an optimized laser focal spot shown in Fig.~\ref{fig:optimization}(C), which decreased the radius containing 50\% of the pulse energy by 13\%. Notably, the integrated flux on the ToF spectrometer increased over the course of the optimization (Fig.~\ref{fig:optimization}(B), third panel), although this parameter was not considered by the optimizer. The proton flux observed on the ToF spectrometer for the optimal burst was 60\% higher than the initial total flux---a substantially larger change than observed for other properties of the proton beam. Increases in flux may have reduced shot-to-shot variation in the measured maximum proton energy, improving the model's confidence in these regions of parameter space and incentivizing continued data acquisition there.

\section{Discussion}
While the optimization process described here showed that the best proton acceleration performance is associated with improved energy containment within the focal spot, earlier experiments studying the impact of laser wavefront shaping on the properties of laser-driven electron beams\cite{Mangles2009, Beaurepaire2015} found that aberrated laser wavefronts can lead to more favorable properties in the resulting electron beam. Similarly, in our previous work on BO of laser-driven proton beams using a tape drive target\cite{Loughran2023} the optimal focal spot was significantly aberrated compared to the manually-optimized focal spot. This difference in performance suggests that the \qty{600}{\nano\meter} thick liquid jet target used here may have been more sensitive to subtle degradations in the laser maximum intensity than the micron-scale tape targets used in Ref.~\citenum{Loughran2023}. The liquid jet target was significantly thinner than the tape, potentially causing the target rear surface (and hence the ion acceleration process) to be more easily perturbed by structured intensity distributions incident on the front surface.

Another recent study also found that higher maximum proton energies were obtained from \unit{\micro\meter}-scale foil targets when the laser wavefront was significantly aberrated\cite{Catrix2025}. This study also sought to optimize maximum proton energy using BO of DFM parameters, but focused on tuning individual DFM actuators rather than the Zernike mode coefficients resulting from multi-actuator adjustments of the DFM surface. This approach required a much higher-dimensional optimization process due to the use of individual DFM actuators as optimization parameters, but the optimization results were dominated by a single actuator: the central ``focus'' actuator. The remaining actuators each made relatively small contributions to the result.

\begin{figure}
    \centering
    \includegraphics[width=\linewidth]{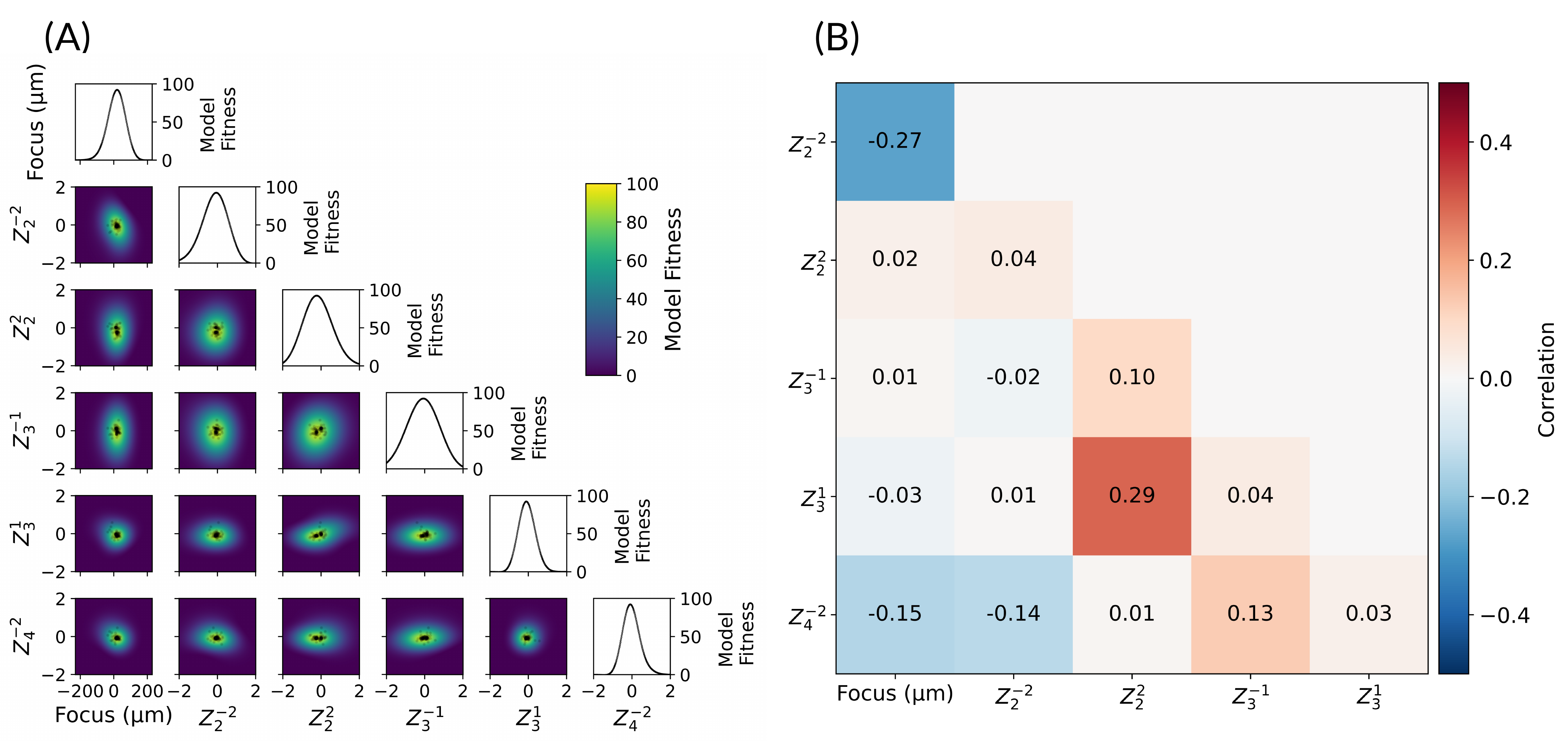}
    \caption{(A): 1D and 2D surfaces of the fitness values predicted by the final optimization model as a function of the Zernike coefficients used as optimization parameters. 1D plots along the upper right indicate the fitness predicted by the model as one parameter is varied, with all other parameters held at the optimal value. 2D contour plots indicate the fitness predicted by the model as two parameters are varied, with all other parameters held at the optimal values. Narrower peaks indicate more sensitivity to changes in the parameter value. 2D surfaces with components along the diagonal (rather than purely vertical or horizontal) directions indicate coupling between the parameters. The points overlaid in black indicate the positions in parameter space where the BO algorithm acquired data. (B): The 2D plots from (A) were fit to a 2D standard normal probability density function (Eq.~\ref{eq:2dgauss}), then the correlation was extracted. This value is indicated for each pairwise combination of parameters, and the matrix element is colored accordingly. Larger magnitudes indicate greater observed correlation between the two parameters.}
    \label{fig:opt_analysis}
\end{figure}

To better understand the parameter sensitivities in our trained model, we used the model to reconstruct 1D and 2D subspaces of expected fitness as a function of the optimization parameters (Fig.~\ref{fig:opt_analysis}(A)). The model observed more sensitivity to variations in focus, horizontal coma ($Z^{1}_{3}$), and oblique second-order astigmatism ($Z^{-2}_{4}$) compared to the other parameters, though only to a modest extent. The parameters used for the optimization described here therefore contributed more evenly than those used for the optimization described in Ref.~\cite{Catrix2025}.

We computed the pairwise correlation between the parameters by fitting the 2D surfaces shown in Fig.~\ref{fig:opt_analysis}(A) to 2D standard normal probability density functions of the form 
\begin{align}\label{eq:2dgauss}
    f(x,y) &= \frac{1}{2\pi\sigma_{X}\sigma_{Y}\sqrt{1-\rho^{2}}}\exp\left(-\frac{1}{2\left[1-\rho^{2}\right]}\left[\left(\frac{x-\mu_{X}}{\sigma_{X}}\right)^{2}\right.\right.\nonumber\\
    &\qquad\left.\left.-2\rho\left(\frac{x-\mu_{X}}{\sigma_{X}}\right)\left(\frac{y-\mu_{Y}}{\sigma_{Y}}\right)+\left(\frac{y-\mu_{Y}}{\sigma_{Y}}\right)^{2}\right]\right),
\end{align} 
where $\mu_{x,y}$ and $\sigma_{x,y}$ are the means and standard deviations of the two parameters, then extracting the correlation $\rho$ between the parameters. The correlations observed by the model indicate that the Zernike coefficients had coupled effects on the maximum energy of the laser-accelerated protons, either due to the highly nonlinear laser-plasma interaction or as a result of inherent couplings between the phase maps of the Zernike coefficients. The largest magnitude of the correlations observed was between horizontal coma and vertical astigmatism ($Z_{3}^{1}$ and $Z_{2}^{2}$, $\rho=0.29$). The correlation between oblique astigmatism ($Z_{2}^{-2}$) and shifts in the focus had nearly the same magnitude. The remaining correlations were significantly lower in magnitude ($\left|\rho\right|\leq0.15$).

\begin{figure}
    \centering
    \includegraphics[width=\linewidth]{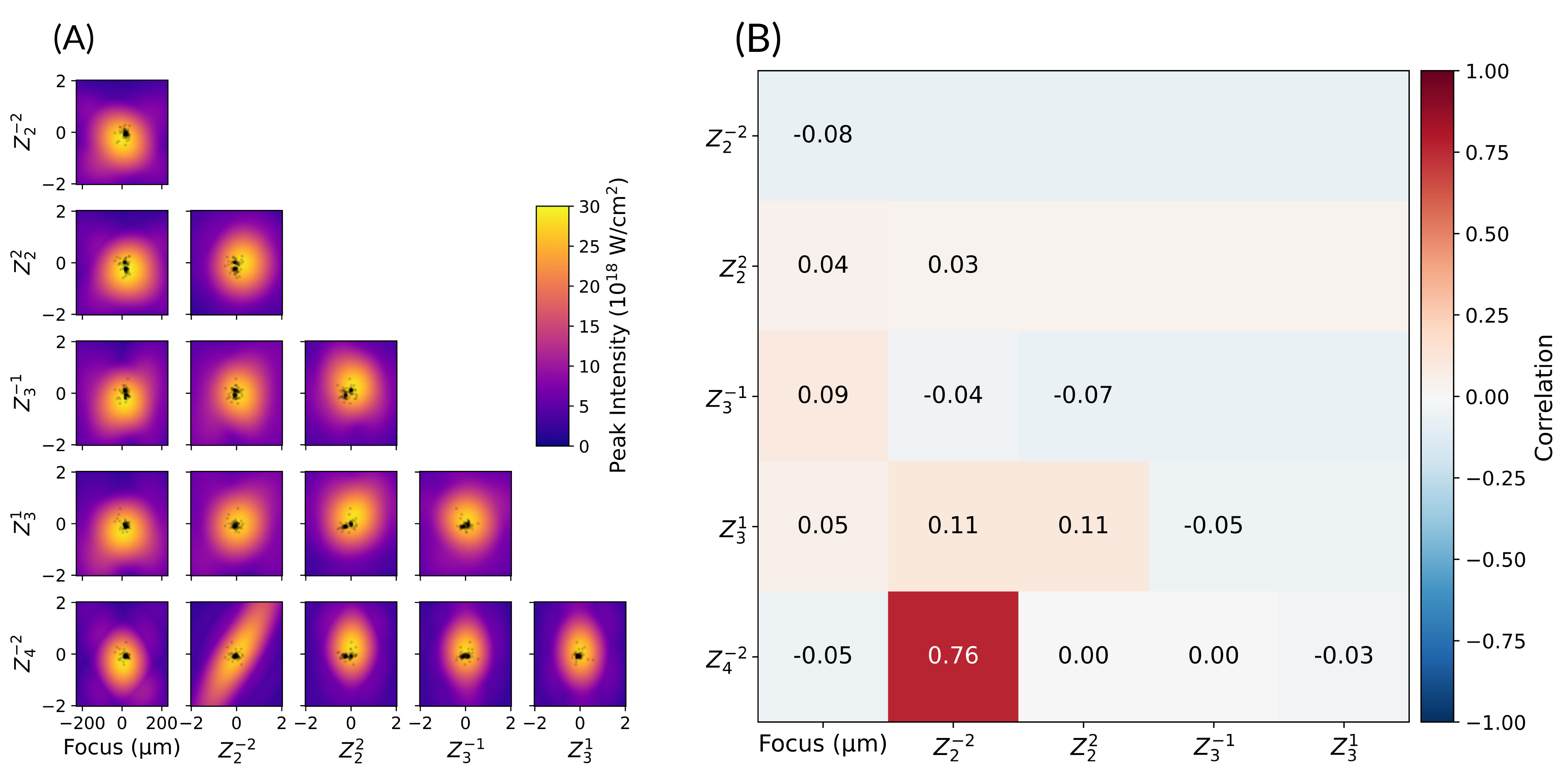}
    \caption{(A) Contour plots depicting the peak laser intensity achieved by laser focal spots reconstructed using the specified values of the Zernike coefficients used in the optimization. For each contour plot, two Zernike coefficients were varied and the remaining four were kept fixed at the optimal values. At each point in the plotted parameter space, a map of the laser far-field intensity distribution was reconstructed based on the Zernike coefficients, then the peak intensity was extracted. The points overlaid in black indicate the positions in parameter space where the BO algorithm acquired data. (B): The 2D plots from (A) were fit to a 2D standard normal probability density function (Eq.~\ref{eq:2dgauss}), then the correlation was extracted. This value is indicated for each pairwise combination of parameters, and the matrix element is colored accordingly. Larger magnitudes indicate greater observed correlation between the two parameters.}
    \label{fig:opt_analysis_discuss}
\end{figure}

One possible explanation for the observed correlations is that they reflect combinations of Zernike modes that preserve the peak intensity in the far field, maintaining higher maximum proton energies despite substantial wavefront distortion. To assess this hypothesis, we reconstructed a series of far-field intensity distributions as a function of the six Zernike coefficients used in the optimization. For each pairwise combination of parameters, two parameters were varied and the remainder were kept fixed at their optima. At each point in parameter space, we reconstructed the far-field intensity distribution and computed the maximum value. The resulting contour plots of peak laser intensity for each combination of Zernike coefficients are shown in Fig.~\ref{fig:opt_analysis_discuss}(A).

The reconstructed peak intensity contour plots show substantial differences from the fitness contours in Fig.~\ref{fig:opt_analysis}(A), indicating a much different set of correlations between the parameters. As quantified in Fig.~\ref{fig:opt_analysis_discuss}(B), the Zernike coefficients considered here almost all showed low levels of correlation ($\left|\rho\right|\leq0.11$) in their effect on the laser peak intensity, including the parameters which were inferred by the optimizer to have the highest correlation in their effect on the proton maximum energy. The only parameters with highly correlated effects on the peak intensity were the oblique astigmatism ($Z_{2}^{-2}$) and the oblique second-order astigmatism ($Z_{4}^{-2}$)---as expected given that these two terms can partially compensate for one another. However, this correlation was not captured by the BO model, perhaps due to its relatively sparse sampling. As shown in Fig.~\ref{fig:opt_analysis_discuss}(A), the optimizer only explored a limited region of the parameter space, instead focusing on the regions where the laser intensity was maximized. The acquisition function used here was intended to optimize the maximum proton energy, rather than map the parameter space, likely explaining the optimizer's preference for remaining near the highest laser intensities. Further, sampling a six-dimensional parameter space with only ${\sim}$100 points inherently means that the space will be relatively sparsely sampled, particularly when the points are highly unevenly distributed.

We therefore conclude that the parameter correlations observed by the BO model cannot be explained simply by their effect on the laser peak intensity. Further study is necessary to determine whether these correlations simply result from the sparse sampling of the parameter space or are real features emerging from complex interactions between the laser wavefront and target plasma. Alternative acquisition functions such as the ``Bayesian exploration'' method\cite{Roussel2024} could be used to map out the parameter space in more detail to facilitate a clearer understanding of the relationship between each Zernike mode coefficient and the maximum proton energy. This work could be supplemented by a series of 2D and 3D numerical simulations with careful adjustments to the laser wavefront to monitor the impact on plasma dynamics and the maximum proton energy. Due to the expense of such a simulation campaign, it may still be more feasible to investigate these effects experimentally, taking advantage of the large datasets available from stable HRR lasers and targets.

\section{Conclusion}
In this work we have presented a variety of measurements of laser-accelerated particle beams produced in high-intensity, high repetition rate laser interactions with a liquid water sheet jet target. Optical probing with an independent laser pulse allowed us to observe the growth of filamentary plasma structures in the first tens of \unit{\pico\second} after the interaction and the initial formation of a laser-driven shock propagating away from the interaction volume. Consistent with previous work, we observed significant improvements in the energy and flux of laser-accelerated protons when p-polarized laser pulses were used. Finally, we performed real-time, closed-loop optimization of the proton maximum energy by tuning the laser wavefront using a deformable mirror. The optimizer improved the energy concentration in the focal spot compared to manual tuning prior to the autonomous optimization experiment, leading to improved proton beam maximum energies. The optimizer inferred correlations between the Zernike mode coefficients used to tune the laser wavefront, but these correlations did not correspond directly to the changes in the far-field peak intensity generated by the same changes to the Zernike coefficients, suggesting that the inferred correlations may have resulted from more subtle shaping of the laser wavefront rather than solely changes to the maximum intensity alone. Further study of the detailed interaction between the laser wavefront and the target front-surface plasma, either via high-resolution experimental parameter scans or numerical simulations, will be necessary for a full exploration of the physics underlying these inferred correlations. 

The use of a highly stable, replenishing target was key to these results, making it possible to collect statistics throughout scans across a number of experimental parameters. The automated optimization work shown here can be readily extended, such as by changing the BO reward function to optimize different experimental observables, such as the proton beam spatial profile or flux, or by adjusting the acquisition function to prioritize parameter space exploration rather than optimization. Alternatively, optimizations could be performed using other parameters than the laser wavefront, such as spectral phase components governing the pulse temporal profile\cite{Mariscal2024} or the energy or delay of an independent probe beam used to pre-expand the target\cite{Tamminga2025}. Liquid sheet targets, with thicknesses that can be adjusted in-situ, may also permit tuning between acceleration regimes that favor higher maximum energies and those that favor increased ion flux by varying the target thickness. Finally, operation with \unit{\peta\watt}-class laser drivers will enable optimization of ion beams with maximum energies of tens of \unit{\mega\electronvolt}. The capabilities demonstrated here will guide the continued maturation of HRR laser-driven ion sources, demonstrating the potential for the same source to both provide insights into fundamental ion acceleration physics and to facilitate real-time optimization of laser-driven proton beams as needed for applications.

\section{Methods}\label{sec:methods}
\subsection{TA2 laser system}
The experiment was performed in the Gemini TA2 target area at the Central Laser Facility (UK). The maximum laser energy of \qty{200}{\milli\joule} was measured on-shot by imaging the near-field of the beam through a dielectric mirror onto a CCD which had been cross-calibrated to a Gentec energy meter prior to the compressor. Compressor and post-compressor beamline throughput was measured at regular intervals and included within the determination of on-target energy. A HASO wavefront sensor coupled to a deformable mirror enabled both measurement and manipulation of the laser spatial phase. The laser compressor was separated from the target chamber using a \qty{0.5}{\milli\meter} thick fused silica pellicle. The full-energy compressed pulse passed through this window, adding a maximum B-integral of 0.2. A sample of the compressed pulse, \qty{5}{\milli\meter} in diameter, passed through an identical pellicle and was then attenuated and passed through an additional \qty{2}{\milli\meter} thick fused silica vacuum window. The temporal profile of the pulse sample was then characterized using a SPIDER\cite{Iaconis1998} and numerically back-propagated through \qty{2}{\milli\meter} of fused silica to reconstruct the on-target pulse duration of $75\pm5$~\unit{\femto\second}. The pulse duration, group delay dispersion (GDD), and third-order dispersion (TOD) were tuned during the experiment using a Fastlite Dazzler acousto-optic programmable filter. The beam was focused in the interaction chamber using an $f/2.5$ off-axis parabola, yielding a peak intensity on target of $3.5\times10^{19}$ \qty[per-mode=symbol]{}{\watt\per\square\centi\meter} (normalized vector potential $a_{0}\approx4$). The beam was incident onto the target at a \qty{30}{\degree} angle from the target front surface normal. The polarization was varied using a full aperture quarter- or half-waveplate placed in the beam. The quality of the focal spot was monitored periodically by imaging the attenuated laser beam using a Mitutoyo $50\times$ infinity-corrected apochromatic microscope objective coupled to a turning mirror and camera. Before the experiment the laser contrast was measured to be of order $10^{-8}$ greater than \qty{20}{\pico\second} before the main pulse and of order $10^{-5}$ greater than \qty{5}{\pico\second} before the main pulse.

\subsection{Liquid water sheet jet and vacuum system modifications}
The liquid sheet jet was generated using a tungsten microfluidic nozzle\cite{Treffert2022a}. Water was pumped into the nozzle from outside the vacuum chamber using a high pressure liquid chromatography (HPLC) pump, then injected into vacuum through a $100\times25$~\unit{\square\micro\meter} converging aperture at an average flow rate of \qty[per-mode=symbol]{6}{\milli\liter\per\minute}. The nozzle produced a liquid sheet approximately \qty{3}{\milli\meter} long in the vertical dimension and \qty{0.5}{\milli\meter} wide. The liquid flow speed of approximately \qty[per-mode=symbol]{10}{\meter\per\second} ensured that the interaction region was completely refreshed before each laser shot. The target thickness was measured using white light interferometry and was determined to be $600\pm100$~\unit{\nano\meter} at the interaction point \qty{2.8}{\milli\meter} below the nozzle. Measurement of the jet position using the probe beam determined that the position jitter of the sheet edges in the plane of the sheet was less than \qty{5}{\micro\meter}. It is expected that the jitter in the plane of the sheet was significantly larger than the jitter in the perpendicular direction due to the geometry of the liquid sheet formation. The sheet surface therefore remained well within the \qty{15}{\micro\meter} Rayleigh range of the high-power laser.

To reduce the load on the vacuum pumps, the liquid jet was intercepted below the interaction point by a differentially-pumped in-vacuum catcher with a \diameter\qty{500}{\micro\meter} aperture. The catcher was heated to a temperature of approximately \qty{470}{\kelvin}, causing the liquid jet to vaporize inside the catcher body after passing through the aperture. The water vapor was evacuated using a gravity-assisted liquid trap attached to a scroll pump. The vacuum chamber maintained a pressure of \qty{0.1}{\milli\bar} during liquid jet operation.

\subsection{Experimental diagnostics}
\subsubsection{Optical probe}
A probe beam was generated using the laser frontend ($<1$~\unit{\milli\joule}) and was independently compressed to a pulse duration ${\sim}50$~\unit{\femto\second}. The probe was incident on the target at an angle of approximately \qty{20}{\degree} from the front surface normal, \qty{50}{\degree} from the main beam axis. A motorized delay stage provided up to \qty{1.6}{\nano\second} of delay between the main laser interaction and the arrival of the probe. The $t=0$ position was determined by observation of interference fringes between the probe and main pulse in images of the probe beam obtained during a very fine timing scan of the probe. This observation was coincident with the earliest observations of perturbation of the sheet by the arrival of the main beam. All images presented in Fig.~\ref{fig:probefig} have been normalized by an image of the jet collected without any drive laser irradiation.

\subsubsection{Proton beam diagnostics}
A \qty{2}{\square\milli\meter} diamond detector operating in time-of-flight mode measured the ion energy spectrum. The detector was placed \qty{357}{\milli\meter} from the target with a line of sight at an angle of 3$^\circ$ relative to the rear-surface target normal. No filters were included ahead of the detector, so the contribution of heavier ions was separated from the proton signal by the arrival time at the detector. Previous work has shown that the maximum energy of the heavy ions per nucleon is typically less than the maximum proton energy\cite{Hegelich2002,Carroll2010}, meaning that the proton maximum energy cutoff we observed should not be sensitive to heavy ion contributions. Further, as we do not observe a large secondary peak from the arrival of low-energy heavy ions, we estimate that the high-energy heavy ion flux does not make a significant contribution to the ToF signal we ascribe to laser-accelerated protons.

The spatial profile of the rear-surface proton beam was captured using a \diameter\qty{50}{\milli\meter} ZnS(Ag) scintillator (EJ-440, Eljen Technologies) placed \qty{160}{\milli\meter} from the target, centered on the rear-surface normal. The scintillator was shielded from scattered laser light and low-energy heavy ions using a \qty{12}{\micro\meter} layer of aluminized Mylar, making it most sensitive to \qty{1.1}{\mega\electronvolt} protons over most of its surface. Selected regions were additionally filtered using layers of \qty{10}{\micro\meter} Al to provide maximum sensitivity to \qty{1.5}{\mega\electronvolt}, \qty{1.9}{\mega\electronvolt}, and \qty{2.2}{\mega\electronvolt} protons in regions covered by an additional \qty{10}{\micro\meter}, \qty{20}{\micro\meter}, and \qty{30}{\micro\meter} of Al, respectively. The energy deposition per proton as a function of proton energy was determined using FLUKA\cite{Ballarini2024} simulations.

The scintillator was absolutely calibrated by conversion to a slotted radiochromic film (RCF) stack consisting of one layer of Gafchromic HD-V2 and one layer of EBT3 behind a \qty{12}{\micro\meter} thick aluminized Mylar shield. The slots allowed samples of the proton beam to propagate to the scintillator while the rest of the beam was captured by the RCF stack. The HD-V2 film was scanned and converted to a dose map using the procedure described in Ref.~\citenum{Greenwood2022}, and the resulting map was compared to the camera signal to determine an absolute calibration.

The proton beams generated in these interactions exhibited remarkably low divergence on the order of \qty{1}{\degree}\cite{Streeter2025}, making it essential to track the ion beam pointing on-shot. Since the ToF spectrometer sampled only a small solid angle, fluctuations in the proton beam pointing of only a few degrees could potentially cause substantial changes to the measured proton energy spectrum. This sensitivity stands in contrast to the relative insensitivity of such diagnostics to the pointing of proton beams accelerated via sheath acceleration from conventional targets, which exhibit much larger typical divergences. For all proton spectra and maximum energies shown in this work, it was necessary to verify that the proton beams were pointed towards the ToF spectrometer using the proton spatial profiler in order to ensure that the energy spectra shown were sampled from similar regions of the beam.

\subsubsection{Electron diagnostic}
To record the energy spectrum of electrons escaping from the interaction, a permanent magnet-based in-vacuum electron spectrometer was placed along the laser propagation direction at a distance of \qty{350}{\milli\meter} from the interaction. Electrons passed through a $0.5\times4$~\unit{\square\milli\meter} slit (acceptance of \qty{35}{\micro\steradian}) and were deflected onto a Kodak Lanex Regular scintillating screen using a permanent magnet with a peak field of approximately \qty{0.15}{\tesla}. The light emitted from the screen was captured by a CCD camera on each laser shot. The energy dispersion of the spectrometer was determined with particle tracing simulations using the measured magnetic field distribution (obtained with a Hall probe). The minimum observable electron energy was approximately \qty{1}{\mega\electronvolt}.

\subsection{Bayesian optimization algorithm}
The algorithm used for the Bayesian optimization demonstrated here was a home-built algorithm implemented in scikit-learn which was substantially identical to the algorithm used in Ref.~\citenum{Shalloo2020}. The acquisition function used was a variation of the expected improvement acquisition function that accounted for the proportion of uncertainty attributable to model error as opposed to measurement noise. The kernel was a radial basis function added to an additional white noise function. Hyperparameters were optimized in each iteration of the model via maximum likelihood estimation. For further details, see the Methods section of Ref.~\citenum{Shalloo2020}. In contrast to Ref.~\citenum{Shalloo2020}, we implemented a different objective function aimed at optimizing the maximum energy observed in the ToF detector.

\begin{acknowledgments}
Special thanks goes to the staff at the Central Laser Facility who provided laser operational support, mechanical and electrical support, and computational and administrative support. S.H.G., G.D.G., C.P., M.G., C.C., and F.T. acknowledge support from the U.S. DOE Office of Science, Fusion Energy Sciences under FWP No.~100182, and in part from the NSF under grant No.~1632708 and PHY-2308860. G.D.G. acknowledges support from the DOE NNSA SSGF program under DE-NA0003960. M.J.V.S. acknowledges support from the Royal Society URF-R1221874. A.G.R.T. and S.D. acknowledge support from the U.S. DOE Grant No. DE-SC0016804 and U.S. Air Force Office of Scientific Research Grant No. FA9550-19-1-0072 and U.S. Department of Energy NNSA Center of Excellence under cooperative agreement number DE-NA0004146. Z.N., O.E., G.H., and N.X. acknowledge support from the JAI, STFC grant No. ST/P002021/1 and ST/V001639/1. P.McK., R.G. and M.K. acknowledge support from EPSRC grant number EP/R006202/1. C.A.J.P. acknowledges support from EPSRC grant number EP/Y001737/1. B.L. acknowledges support from UK XFEL Physical Sciences Hub under agreement No. S2-2020-00020-8457. L.G. acknowledges support by the M\v{S}MT \v{C}R Project No. LQ1606 and by the project CZ.02.1.01/0.0/0.0/16\_019/0000789.
\end{acknowledgments}

\bibliography{GDGlenn_2025_References}

\end{document}